\documentclass[reprint,prl,aps]{revtex4-1}
\newcommand{\be}{\begin{equation}}
\newcommand{\ee}{\end{equation}}
\newcommand{\ben}{\begin{eqnarray}}
\newcommand{\een}{\end{eqnarray}}
\newcommand{\half}{\mbox{$\case{1}{2}$}}
\newcommand{\quart}{\mbox{$\case{1}{4}$}}
\begin{document}
\title{Generalized non-additive entropies and quantum entanglement}
\author{N. Canosa, R. Rossignoli}
\address{Departamento de F\'{\i}sica, Universidad Nacional de La Plata,
C.C.\ 67, 1900 La Plata, Argentina} 

\begin{abstract}
We examine the inference of quantum density operators from incomplete
information by means of the maximization of general non-additive entropic
forms. Extended thermodynamic relations are given. When applied to a bipartite
spin $\half$ system, the formalism allows to avoid fake entanglement for data
based on the Bell-CHSH observable, and, in general, on any set of Bell
constraints. Particular results obtained with the Tsallis entropy and with an
introduced exponential entropic form are also discussed.\hfill\break
%%%%%%%%%%%%%%%%%%%%%%%%%%%%%%%%%%%%%%%%%%%%%%%%%%%%%%%%%%%%%%%%%%%%%%%%%%%%%%%
Pacs:  05.30.-d, 03.67-a, 03.65.Ud\hfill\break
\end{abstract}
\maketitle

The relation between two fundamental concepts, entropy and quantum
entanglement, has recently aroused great interest in quantum information theory
\cite{HHH.99,R.99,AR.99,TLB.01,GG.01,HH.96,CA.97,BZ.99}. A system composed of
two subsystems $A$ and $B$ is said to be {\it unentangled} or {\it separable},
if and only if the density operator $\rho$ can be written as a convex
combination of uncorrelated densities, i.e.,
$\rho=\sum_jq_j\rho^A_j\otimes\rho^B_j$, with $q_j\geq 0$. 
In this case the system admits a local description in terms of hidden variables. 
Otherwise, it is said to be {\it entangled} or {\it inseparable}. The system becomes 
then suitable, in principle, for applications like quantum cryptography
\cite{Ek.91} and teleportation \cite{BBCJPW.93}.

When the available information about the system is {\it incomplete}, consisting
for instance of the expectation values of a reduced set of observables, one
faces the problem of first determining if entanglement is actually implied by
the data, and then selecting the most probable or representative density
operator compatible with these data. An ideal inference scheme in this scenario
should then a) avoid {\it fake entanglement} \cite{HHH.99}, i.e., should not
yield an entangled density if there is a separable density that reproduces the
data, b) be {\it least biased}, in the sense that {\it some} measure of lack of
information is maximized, and c) be simple enough to be readily applied. As
shown in \cite{HHH.99}, the standard approach based on the direct maximization
of the {\it  von Neumann} entropy $S=-{\rm Tr}\,\rho\ln\rho$, does not comply
with a) already for two spin $\half$ systems. The essential reason is that this
entropy is not a good entanglement indicator \cite{GG.01,CA.97,BZ.99}, even in
those cases where entanglement is fully determined by the eigenvalues of
$\rho$. A solution was also provided in \cite{HHH.99}: one should first
determine the set of densities that minimize entanglement, and then maximize
entropy within this set. Although certainly rigorous, this procedure is not
easy to implement in general, and departs conceptually from a more basic
approach based on the maximization of a single information measure.

As is well known, the von Neumann entropy is based on the Shannon information
measure, which is the unique one satisfying the four Khinchin axioms
\cite{BS.93}. However, if the fourth axiom, which is concerned with additivity,
is lifted, other information measures become feasible. The most famous recent
example is the {\it Tsallis} entropy \cite{T.88}, which has been applied to a
wide range of phenomena characterized by non-extensivity \cite{T.95}, including
recently the problem of quantum entanglement \cite{AR.99,TLB.01}.

The aim of this work is first to discuss more general non-additive entropic
forms, based on {\it arbitrary} concave functions, and the ensuing densities
that maximize these forms subject to given constraints. Though sharing many
features with the usual von Neumann based statistics, the extended formalism
opens new possibilities, in particular that of approaching, for certain
functions, densities whose {\it largest eigenvalue} has the {\it minimum value}
compatible with the available data. For a system of two qubits, this allows to
satisfy previous requirements a), b), c) {\it simultaneously} for any set of
Bell constraints, by means of a {\it single} maximization. In particular, for
information based on the Bell-CHSH observable \cite{HHH.99}, we will show that
fake entanglement can be averted even without including data about the
dispersion (in contrast with \cite{R.99,AR.99}), for a wide class of entropic
functions. As particular cases, we will examine results obtained with the
Tsallis entropy, applied here with {\it standard} expectation values (as
opposed to \cite{AR.99}, where the $q$-expectation values were used), and will
also introduce an exponential entropic function, which will provide fully
analytic results.

Given a density operator $\rho=\sum_ip_i|i\rangle\langle i|$, with $p_i\geq 0$,
${\rm Tr}\,\rho=\sum_ip_i=1$ and the sum running over a complete set of
orthonormal states ($\sum_i|i\rangle \langle i|=I)$, we will consider the
entropic form
 \be
S_f(\rho)\equiv{\rm Tr}f(\rho)=\sum_if(p_i)\,,\label{1}
 \ee
where $f$ is a smooth {\it concave} function ($f''(p)<0$ for $p\in(0,1)$ and
$f(p)$ continuous for $p\in[0,1]$) satisfying $f(0)=f(1)=0$. Although Eq.\
(\ref{1}) is certainly not the most general information measure, it is the most
simple generalization of the von Neumann entropy ($f(p)=-p\ln p$) and comprises
useful extensions. With the exception of additivity, the basic properties of
entropy are satisfied for an arbitrary function $f$ of this form. In
particular, $S_f(\rho)\geq 0$, with $S_f(\rho)=0$ only in the case of a pure
state ($\rho^2=\rho$), its maximum is attained for a uniform distribution
($p_i=1/n$, with $n$ the space dimension),  with the maximum value $nf(1/n)$
increasing with $n$, and it is not affected by states with $p_i=0$. Concavity
of $f$ implies concavity of $S_f(\rho)$ \cite{W.78}, i.e.,
 $S_f(\sum_jq_j\rho_j)\geq \sum_jq_jS_f(\rho_j)$, with $q_j\geq 0$,
$\sum_jq_j=1$, as well as the important property that $S_f(\rho)$ {\it cannot
decrease by dephasing}, i.e., by removing the off-diagonal elements of $\rho$
in any basis of orthonormal states  $|i_o\rangle$:
 \be
S_f(\rho)\leq S_f(\rho_o)\,,\;\;\;\;\;\;\rho_o= \sum_i\langle
i_o|\rho|i_o\rangle |i_o\rangle\langle i_o|\,.\label{2}
 \ee
A sufficient condition for $S_f$ to be sub (super) additive is that $pf''(p)$
be a decreasing (increasing) function of $p$ for $p\in(0,1)$, since in this
case $f(pq)-qf(p)-pf(q) \leq 0$ ($\geq 0$) $\forall \; p,q \in[0,1]$, implying
\cite{RC.99}
 \be
S_f(\rho^A\otimes \rho^B)\,{\leq\atop\geq}\,S_f(\rho^A)+S_f(\rho^B)
 \;\;{\rm if}\;\;(pf''(p))'\,{\leq\atop\geq}\,0\,. \label{crit}
 \ee
Additivity amongst entropies of the form (1) holds only if $(pf''(p))'=0$,
which leads immediately to the Shannon form $f(p)=-k\,p\ln p$,  $k>0$.

Maximization of $S_f(\rho)$ subject to the constraints of $m+1$ expectation
values $\langle O_\alpha\rangle={\rm Tr}\rho O_\alpha$, $\alpha=0,\ldots,m$,
where $O_\alpha$  are linearly independent observables (not necessarily
commuting) and we have set $O_0=I$ to account for normalization
 ($\langle I\rangle=1$), leads to the maximization of
 \be
\bar{S}_f(\rho)=S_f(\rho)-\sum_\alpha\lambda_\alpha \langle
O_\alpha\rangle\label{Sbf}={\rm Tr}[f(\rho)-\rho h],
 \ee
where $h=\sum_\alpha\lambda_\alpha O_\alpha$ and $\lambda_\alpha$ are Lagrange
multipliers. Writing $h=\sum_i h_i|i_h\rangle\langle i_h|$, the maximum is
attained for
 \ben
\rho&=&p(h)=\sum_ip_i|i_h\rangle\langle i_h|\label{Dm}\,,\\ p_i&=&p(h_i)\equiv
\left\{\begin{array}{ll}[f']^{-1}(h_i) &\;\; f'(1)\leq h_i<f'(0)\\
0&\;\;h_i\geq f'(0)\end{array}\right.\,, \label{ph}
 \een
where $[f']^{-1}$ is the inverse of the function $f'$. The cut-off for
 $h_i\geq f'(0)$ obviously arises only when $f'(0)$ is {\it finite}, and
is the main difference with the von Neumann case (where Eq.\ (\ref{ph}) becomes
the exponential distribution $p_i=e^{-(h_i+1)}$, with $h_i\geq -1$).
Nevertheless, due to the concavity of $f$, $p_i$ is always a {\it
non-increasing} function of $h_i$ ($p'(h_i)=1/f''(p_i)<0$ if $f'(1)<h_i<f'(0)$,
and $0$ if $h_i>f'(0)$) that vanishes for $h_i\rightarrow\infty$ $\forall$ $f$.

Eqs.\ (\ref{Dm})--(\ref{ph}) can be easily derived. As
 ${\rm Tr}\,\rho h={\rm Tr}\,\rho_h h$, with
$\rho_h=\sum_i\langle i_h|\rho|i_h\rangle |i_h\rangle\langle i_h|$, Eq.\
(\ref{2}) implies $\bar{S}_f(\rho)\leq \bar{S}_f(\rho_h)$. The optimum density
satisfies then $[\rho,h]=0$, so that $\rho_h=\rho$ and
$\bar{S}_f(\rho)=\sum_if(p_i)-p_ih_i$. Due to the concavity of $f$, this will
have  a {\it unique} maximum for $p_i\in[0,1]$, determined by $f'(p_i)=h_i$ if
$f'(1)<h_i<f'(0)$, or otherwise located at one of the borders, which leads to
Eq.\ (\ref{ph}) (for a non-standard normalization $\langle I\rangle>1$,  an
upper cutoff $p_i=1$ if $h_i<f'(1)$ would also apply).

Considering  $\bar{S}_f(\rho)$ and $S_f(\rho)$ at the maximum (\ref{Dm}) as
functions of $\bbox{\lambda}\equiv(\lambda_0,\ldots,\lambda_m)$ and
$\bar{\bbox{O}}\equiv(\langle O_0\rangle,\ldots,\langle O_m\rangle)$
respectively, we obtain the thermodynamic relationships
 \ben
\frac{\partial\bar{S}_f(\bbox{\lambda})}{\partial\lambda_\alpha}&=&
 -\langle O_\alpha\rangle\,, \;\;\;\;\;\;\;
\frac{\partial^2\bar{S}_f(\bbox{\lambda})}
{\partial\lambda_\alpha\partial\lambda_{\beta}}=A_{\alpha\beta}\,,
\label{Sfb1}\\
\frac{\partial S_f(\bbox{\bar{O}})}{\partial\langle O_\alpha\rangle}
&=&\lambda_\alpha\,,\;\;\;\;\;\;\;\;\;\;\;\;
\frac{\partial^2{S}_f(\bbox{\bar{O}})}
 {\partial\langle O_\alpha\rangle\partial\langle O_{\beta}\rangle}=
-(A^{-1})_{\alpha\beta}\,,\label{Sfb2}\\
 A_{\alpha\beta}&=& \sum_{i,j}\langle i_h|O_\alpha|j_h\rangle
 \langle j_h|O_{\beta}|i_h\rangle C_{ij}\,,\label{A}\\
 C_{ij}&=&-\delta_{ij}p'(h_i)+(1-\delta_{ij}) \frac{p_j-p_i}{h_i-h_j}\geq
0\label{C}\,.
 \een
Only the {\it second} derivatives in (\ref{Sfb1})--(\ref{Sfb2}) depend
explicitly on $f$, through the first term in (\ref{C}). As  $C_{ij}\geq 0$,
$A_{\alpha\alpha}\geq 0$. Moreover, all eigenvalues of $A$ are {\it
non-negative}, i.e., $A_\alpha=\sum_{i,j}
 |\langle i_h|Q_\alpha|j_h\rangle^2|C_{ij}$, with $Q_\alpha=\sum_{\beta}
U_{\beta\alpha}O_{\beta}$ and $U$ defined by $[U^{\rm tr}AU]_{\alpha\beta}
=A_\alpha\delta_{\alpha\beta}$. Hence $\bar{S}_f$ is a {\it convex} function of
$\bbox{\lambda}$, whereas $S_f$ a {\it concave} function of $\bbox{\bar{O}}$,
as in the von Neumann  case. If $[O_\alpha,O_\beta]=0$ $\forall$
$\alpha,\beta$, Eq.\ (\ref{A}) becomes
 $A_{\alpha\beta}=-{\rm Tr}\,\rho'\,O_\alpha O_\beta$, with $\rho'\equiv
\sum_ip'(h_i)|i_h\rangle\langle i_h|$ (for $f(p)=-p\ln p$, $p'(h_i)=-p_i$ and
$\rho'=-\rho$).

We will be here interested in functions of the form
 \be
f(p)=k(p-g_q(p))\,,\label{fgq}
 \ee
where $k>0$ and $g_q(p)$ is a convex ($g''_q(p)>0$) {\it increasing}
function satisfying $g_q(0)=0$, $g_q(1)=1$, and
 \be
\lim_{q\rightarrow\infty} g_q(p_i)/g_q(p_j)=0\;\;{\rm if}\;\;
p_i<p_j\,.\label{des}
 \ee
Hence, for sufficiently large $q$ (and finite dimension $n$)
\[S_f(\rho)=k[1-{\rm Tr}\,g_q(\rho)]\approx k(1-n_Mg_q(p_M))\,, \]
where $p_M$ is the largest eigenvalue of $\rho$ and $n_M$ its multiplicity. The
density that maximizes $S_f(\rho)$ (i.e., minimizes ${\rm Tr}\,g_q(\rho)$)
subject to a given set of constraints, will then possess, for
$q\rightarrow\infty$, {\it the minimum largest eigenvalue $p_M$} compatible
with the available data, as in this case $n_Mg_q(p_M)$ is {\it minimum}. This
property may in fact be  fulfilled {\it already for finite values of $q$}
(i.e., typically $q>q_c$) in some cases, as will be seen below.

Similarly, maximization of the entropy associated with
 \be
\tilde{f}(p)=f(1-p)=k(1-p-g_q(1-p))\label{ft}\,,
 \ee
which is also concave and satisfies $\tilde{f}(0)=\tilde{f}(1)=0$, leads to a
density which possesses, for $q\rightarrow\infty$,  the {\it maximum smallest
eigenvalue} compatible with the available data. In this limit
$g_q(1-p_i)/g_q(1-p_j)\rightarrow 0$ if $p_i>p_j$, so that
$S_{\tilde{f}}(\rho)\approx k[n-1-n_mg_q(1-p_m)]$ for large $q$, with $p_m$ the
{\it smallest} eigenvalue and $n_m$ its multiplicity. This is maximum if $p_m$
is maximum.

As a particular example, we have in first term
 \be
f(p)=(p-p^{q})/(q-1)\,,\;\;\;\;q>0\,,\label{q}
 \ee
which is concave for $q>0$, approaches $-p\ln p$ for $q\rightarrow 1$ and is of
the form (\ref{fgq}) for $q>1$, satisfying (\ref{des}). In this case,
$S_f(\rho)=(1-{\rm Tr}\,\rho^q)/(q-1)$ is the {\it Tsallis entropy}, which is
sub (super)-additive for $q>1$ ($q<1$), in agreement with Eq.\ (\ref{crit})
($(pf''(p))'=q(1-q)p^{q-2}$). The $q=2$ case is particularly relevant, since
maximization of $S_f(\rho)$ becomes  equivalent to the minimization of
 ${\rm Tr}\,\rho^2=(\sum_{i<j}(p_i-p_j)^2+1)/n$, i.e., to a {\it least squares
condition } for the probabilities or their differences. For $q=2$, $S_f(\rho)$
also coincides with the information measure of ref.\ \cite{BZ.99}. Eq.\
(\ref{ph}) becomes, setting $h_c=f'(0)$,
\[p_i=\{[1-(q-1)h_i]/q\}^{1/(q-1)}\,,\;\;-1\leq h_i<h_c\]
and $p_i=0$ if $h_i\geq h_c$, with $h_c=\case{1}{q-1}$ ($\infty$) if $q>1$
($q<1$).

Another example is the exponential function
 \be
f(p)=q^{-1}[p-(e^{qp}-1)/(e^{q}-1)]\,,\label{feq}
 \ee
which is concave for any $q$, satisfies $f(0)=f(1)=0$ and approaches $\half
p(1-p)$ for $q\rightarrow 0$, i.e., proportional to the Tsallis case $q=2$. For
$q>0$, it is of the form (\ref{fgq}) and fulfills Eq.\ (\ref{des}). Moreover,
$f_{-q}(p)=f_{q}(1-p)$. As $(pf''(p))'=e^{qp}(1+qp)q/(1-e^q)$, $S_f(\rho)$ is
{\it sub-additive} for $q\geq -1$. Eq.\ (\ref{ph}) becomes
\[p_i=q^{-1}\ln[1-(e^q-1)(h_i-h_c)]\,,\;-1\leq h_i-h_c<0\]
and $p_i=0$ if $h_i\geq h_c$, with $h_c=\case{1}{q}-\case{1}{e^q-1}>0$.

{\it Application}. Let us consider now a bipartite spin $\half$ system.
Starting from the basic unentangled states $|\!\!\uparrow\uparrow\rangle$,
$|\!\!\uparrow\downarrow\rangle$, $|\!\!\downarrow\uparrow\rangle$,
$|\!\!\downarrow\downarrow\rangle$, the  Bell basis is formed by the fully
entangled orthonormal states
 $|\Psi^\pm\rangle=(|\!\!\uparrow\downarrow\rangle\pm
 |\!\!\downarrow\uparrow\rangle)/\sqrt{2}$,
 $|\Phi^\pm\rangle=(|\!\!\uparrow\uparrow\rangle\pm
 |\!\!\downarrow\downarrow\rangle)/\sqrt{2}$
($|\Psi^-\rangle$ is the singlet while $|\Psi^+\rangle$, $|\Phi^\pm\rangle$ are
spin $1$ states with $\langle \bbox{S}\rangle=0$). We will consider here {\it
Bell constraints} \cite{HHH.99}, i.e., mean values of observables which are
{\it diagonal} in the Bell basis. Let us first examine the case of ref.\
\cite{HHH.99}, where the available information is the expectation value of the
(scaled) Bell-CHSH observable
 \be
B=|\Phi^+\rangle\langle\Phi^+|-|\Psi^-\rangle\langle\Psi^-|\,.
 \ee
According to Eq.\ (\ref{Dm}), the density $\rho$ that satisfies
 \be
{\rm Tr}\,\rho=1,\;\;{\rm Tr}\,\rho B=b\,,\;\;\;|b|\leq 1\,,\label{cb}
 \ee
and maximizes (\ref{1}) is of the form 
 \ben
\rho=p(\lambda_0 I+\lambda_1 B)&=&p_0
(|\Psi^+\rangle\langle\Psi^+|+|\Phi^-\rangle\langle\Phi^-|)+\nonumber\\
&&p_+|\Phi^+\rangle\langle\Phi^+|+p_-|\Psi^-\rangle\langle\Psi^-|\,, \label{Ds}
 \een
where  $p_0=p(\lambda_0)$, $p_\pm=p(\lambda_0\pm\lambda_1)$. The constraints
(\ref{cb}) become just $2p_0\!+\!p_+\!+\!p_-=1$, $p_+\!-\!p_-=b$. We may
consider $b\geq 0$, in which case $\lambda_1\leq 0$ and $p_+\geq p_0\geq p_-$,
since for $b\rightarrow -b$, $\lambda_1\rightarrow-\lambda_1$ and
$p_{\pm}\rightarrow p_{\mp}$.

If $f'(1)<\lambda_0\pm\lambda_1<f'(0)$, there is no cut-off and the constraints
lead to the single equation
 \be
f'(p_+)+f'(p_+-b)-2f'({\case{1+b}{2}}-p_+)=0\,,\;\;|b|<b_c\label{eq}
 \ee
which determines $p_+$, and hence $p_-=p_+-b$, $p_0=\half(1+b)-p_+$, for a
given $f$. If $f'(0)$ is {\it finite}, a root of Eq.\ (\ref{eq}) for
$p_+\in[b,\half(1+b)]$ will exist only if $|b|<b_c$, with $b_c$ the root of
 \be
f'(b_c)+f'(0)-2f'(\case{1-b_c}{2})=0\,.\label{bc}
 \ee
Eq.\ (\ref{bc}) is equivalent to $f' (0)=\lambda_0-\lambda_1$, and determines
the onset of the cut-off for $p_-$. It has a single root
$b_c\in[\case{1}{3},1]$, with $b_c\rightarrow 1$ if $f'(0)\rightarrow\infty$.
For $b>b_c$, $\lambda_0-\lambda_1>f'(0)$ and we obtain the solution
 \be
p_+=b\,,\;\;\;p_-=0\,,\;\;\;p_0=\half(1-b)\,,\;\;\;b_c\leq b\leq
1\,.\label{eq2}
 \ee
Eqs.\ (\ref{eq})--(\ref{eq2}) become apparent from the entropy
 \be
S_f(\rho)=f(p_+)+f(p_+-b)+2f(\case{1+b}{2}-p_+)\label{Sb}\,.
 \ee
For fixed $b\geq 0$, Eq.\ (\ref{Sb}) is a concave function of $p_+$ for
$p_+\in[b,\half(1+b)]$, with its maximum located within the interval if
$|b|<b_c$, being then determined by (\ref{eq}), and at the left border if
$b_c\leq b\leq 1$, leading to (\ref{eq2}). At the maximum, $\lambda_1=\partial
S_f(\rho)/\partial b=f'(p_+)-f'(p_0)$ in both cases, with $\lambda_0=f'(p_0)$.
Eqs.\ (\ref{eq})--(\ref{eq2}) imply that $p_+$ is an {\it increasing} function
of $b$ for $b>0$.

For $b\rightarrow 0$, Eq.\ (\ref{eq}) leads to
 \be
p_{+}=\quart(1+2b+\gamma b^2)+O(b^4),\;\;
\gamma=-\quart\frac{f'''(1/4)}{f''(1/4)}\,,\label{bs}
 \ee
where  $\gamma<1$ ($>1$) if $S_f$ is sub (super) additive and satisfies Eq.\
(\ref{crit}). Hence, for $b=0$ we obtain the uniform distribution
$p_+=p_-=p_0=\quart$ for any $f$.  On the other hand, for $b\rightarrow 1$,
$p_+\rightarrow 1$ and $\rho\rightarrow|\Phi^+\rangle\langle\Phi^+|$.

The important question that now arises is whether for a given $f$, the previous
scheme gives fake entanglement. A density $\rho$ diagonal in the Bell basis is
unentangled if and only if its largest eigenvalue is less than $\half$
\cite{HH.96}. The density of the form (\ref{Ds}) that complies with (\ref{cb})
and possesses {\it the minimum largest eigenvalue} corresponds to
 \ben
\begin{array}{ll}p_+=p_0=\quart(1+b),\;\;p_-=\quart(1-3b),\;&0\leq b\leq
\case{1}{3}\\p_+=b,\;\;p_-=0,\;\;p_0=\half(1-b)\,,\;&\case{1}{3}\leq b\leq 1
\end{array}\label{me}
 \een
where $p_+\geq p_0\geq p_-$. Unentangled solutions are then feasible only if
$b<\half$.  Note also that for Bell constraints, entanglement is minimized by
densities which are diagonal in the Bell basis\cite{HHH.99}, so that no
unentangled density of any form complying with (\ref{cb}) exists for $b>\half$.
It is now seen from (\ref{eq2}) that when $f'(0)$ is {\it finite}, the maximum
entropy density {\it coincides with} (\ref{me}) for $b>b_c>\case{1}{3}$. Hence,
as $p_+$ is an increasing function of $b$, {\it fake entanglement will be
avoided for those} $f$ {\it for which} $b_c\leq\half$.

{\it Particular solutions}. In the von Neumann case, Eq.\ (\ref{eq}) yields
$p_+=\quart(1+b)^2$, with $b_c=1$, in agreement with ref.\ \cite{HHH.99} and
Eq.\ (\ref{bs}) ($\gamma=1$). Fake entanglement occurs for
$\sqrt{2}-1<b<\half$.

In the  {\it Tsallis} case (\ref{q}),  $f'(0)$ is finite for $q>1$ and Eq.\
(\ref{bc}) leads to
 \be
b_c=[1+2^{1-1/(q-1)}]^{-1},\;\;\;q>1\,,\label{bcq}
 \ee
which is  a {\it decreasing} function of $q$ satisfying $b_c\leq \half$ for
$q\geq 2$. Hence, {\it fake entanglement will be avoided} $\forall$ $q\geq 2$.
For $q=2$, the solution of Eq.\ (\ref{eq}) is specially simple,
 \be
p_+=\quart(1+2b)\,,\;\;0\leq b<\half,\;\;\;\;p_+=b\,,\;\;\half\leq b\leq 1
\label{q2}
 \ee
which is in agreement with (\ref{bs}) ($\gamma=2-q$) and represents the
solution of {\it minimum squares}. The onset of entanglement occurs here
exactly at $b=b_c$.

Although a simple analytic solution of (\ref{eq}) for arbitrary $q$ is not
feasible, it is easy to verify that Eq.\ (\ref{me}) {\it is obtained for}
$q\rightarrow\infty$ $\forall$ $b$. In this limit, $b_c\rightarrow
\case{1}{3}$, while Eq.\ (\ref{eq}) yields, for large $q$, $p_+\approx
\half(1+b)[1+2^{-1/(q-1)}]^{-1}$, which approaches $\quart(1+b)$ for
$q\rightarrow\infty$.

For the exponential function (\ref{feq}), the solution of Eq.\ (\ref{eq}) is
{\it analytic for any} $q$,
 \ben
p_+&=&\quart(1+2b)-\case{1}{2q}\ln\cosh(\half bq) \,,\;\;0\leq
b<b_c\label{qex}\\
b_c&=&\case{1}{3}+\case{2}{q}\ln[\beta_q-\case{e^{-q/3}}{3\beta_q}]\,,
\;\;\beta_q=[1+\sqrt{1+\case{e^{-q}}{27}}]^{1/3}\,,\nonumber
 \een
with $p_+=b$ for $b\geq b_c$. For $q\rightarrow\infty$,
$b_c\rightarrow\case{1}{3}$ and (\ref{qex}) {\it leads immediately to the
solution with the minimum largest eigenvalue,}  Eq.\ (\ref{me}). Again, $b_c$
is a {\it decreasing} function of $q$, with $b_c<\half$ for $q>0$, so that {\it
fake entanglement is here avoided} $\forall$ $q>0$. For $q\rightarrow 0$,
$b_c\rightarrow\half$  and Eq.\ (\ref{qex}) reduces to (\ref{q2}). For
$b\rightarrow 0$, $p_+\approx \quart(1+2b-\quart qb^2)$, in agreement with
(\ref{bs}). Finally, for $q\rightarrow-\infty$, $b_c\rightarrow 1$ and
$p_+\rightarrow\quart(1+3b)$, with $p_-=p_0=\quart(1-b)$, which is the solution
with {\it the maximum smallest eigenvalue of $\rho$}. This gives fake
entanglement for $b\in[\case{1}{3},\case{1}{2}]$, the {\it maximum interval}
for maximum entropy densities, as $p_+$ is in this case maximum.

{\it Inclusion of the dispersion}. If the dispersion of $B$ is also provided
\cite{R.99}, through the expectation value of
$B^2=|\Phi^+\rangle\langle\Phi^+|+|\Psi^-\rangle\langle\Psi^-|$,  the final
maximum entropy density is actually {\it independent of the choice of} $f$. In
this case $\rho=p(\lambda_0+\lambda_1 B+\lambda_2B^2)$ is also of the form
(\ref{Ds}), with $p_\pm=p(\lambda_0\pm\lambda_{1}+\lambda_{2})$,
$p_0=p(\lambda_0)$, which are {\it completely determined} by the constraints,
i.e., $p_{\pm}=\half(b_2\pm b)$, $p_0=\half(1\!-\!b_2)$, where $b_2={\rm
Tr}\rho B^2=p_+\!+\!p_-$. The only role played here by maximum entropy is to
impose a density of the form (\ref{Ds}), which holds for {\it any} $f$, and
fake entanglement is then always avoided. Note also that when only $b$ is
given, the solution (\ref{eq2}) implies {\it minimum dispersion}, as
$b_2=2p_-+b$ is minimum (see also ref.\ \cite{R.99}).

{\it General Bell constraints}. Higher values of $q$, or even the limit
$q\rightarrow\infty$, may be required in general to avoid fake entanglement.
For example, if $B_\alpha=|\Phi^+\rangle\langle
\Phi^+|-\alpha|\Psi^-\rangle\langle\Psi^-|$, with $\alpha>0$ and ${\rm
Tr}\,\rho B_\alpha=b\geq 0$, unentangled densities complying with these data
are again feasible only if $b\leq\half$. However, in the maximum entropy
density, $b_c$ will {\it increase} as $\alpha$ decreases below $1$, with
$b_c\rightarrow 1$ for $\alpha\rightarrow 0$. This implies a higher threshold
to avoid fake entanglement, namely $q>1+\log_2(1+\alpha^{-1}) $ in the Tsallis
case (\ref{q}) and $q>-4\ln(\alpha)$ in (\ref{feq}). For $\alpha\rightarrow 0$,
both values become $\infty$, although the interval of fake entanglement {\it
vanishes}, approaching $[\half-\case{1}{6}\alpha,\half]$ if $b_c>\half$ (for
$\alpha=0$, $p_+=b$, $p_-=p_0=\case{1}{3}(1-b)$ $\forall$ $f$, becoming $\rho$
a Werner state, and fake entanglement does not occur).

The present arguments are valid for any type of Bell constraints. In this case,
densities of minimum entanglement, as measured by the entanglement of formation
$E_F(\rho)$ \cite{BDiV.96},  are {\it diagonal} in the Bell basis
\cite{HHH.99}, and possess the {\it minimum largest eigenvalue $p_M$}
compatible with the available data if $p_M>\case{1}{2}$. This is so because
$E_F(\rho)$ is an {\it increasing} function of the {\it concurrence} $C(\rho)$,
which for a system of two qubits reduces to  $2p_M-1$ if $p_M>\case{1}{2}$ (and
0 otherwise) when $\rho$ is diagonal in the Bell basis \cite{W.98,GG.01}.
Maximum entropy densities constructed with functions satisfying
(\ref{fgq})--(\ref{des}) will then possess {\it minimum entanglement} for
$q\rightarrow\infty$, although in some cases this may hold already for {\it
finite} values of $q$, as seen in the example. For sufficiently large $q$, the
entropies $S_f(\rho)$ will be essentially {\it decreasing} functions of $p_M$,
being then {\it good entanglement indicators} for these densities. This may
also be the case in systems of $n$ qubits for special constraints that lead to
densities diagonal in a basis of fully entangled states (like the GHZ states
used in \cite{DCT.99}), where separability is again favored by low values of
the largest eigenvalue. Further investigations in this direction are in
progress.

In summary, we have examined the use of general non-additive entropic forms for
the inference of quantum density operators. The formalism enables a direct way
to infer least biased densities with minimum entanglement for a system of two
qubits, and hence to avert fake entanglement, when the information consists of
any set of Bell-constraints. It also provides a general framework for the
description of the thermodynamic aspects of entanglement, as well as a more
deep foundation of the success that non-additive entropies like that of Tsallis
may encounter in this type of problems.

NC and RR acknowledge support from CONICET and CIC, respectively, of Argentina.
%%%%%%%%%%%%%%%%%%%%%%%%%%%%%%%%%%%%%%%%%%%%%%%%%%%%%%%%%%%%%%%%%%%%%%%%%%%%%%%

\end{document}